\documentstyle[aps,twocolumn,epsfig]{revtex}

\def\be{\begin{equation}}
\def\ee{\end{equation}}

\def\R{{\sf I\kern-.15em R}}
\def\T{{\sf T\kern-.45em T}}
\def\C{\kern.1em{\raise.47ex\hbox{$\scriptscriptstyle |$}}
             \kern-.40em{\sf C}}
\def\Z{{\sf Z\kern-.45em Z}}

\begin{document}

\title{Topological relaxation of entangled flux lattices: \\ 
Single {\it vs} collective line dynamics}
\author{Ruslan Bikbov$^{1}$ and Sergei Nechaev$^{1,2}$}
\address{$^{1}$ L D Landau Institute for Theoretical Physics, 117940,
Moscow, Russia \\ $^{2}$LPTMS, Universit\'e Paris Sud, 91405 Orsay Cedex,
France}

\maketitle

\begin{abstract}
A symbolic language allowing to solve statistical problems for the systems with
nonabelian braid--like topology in 2+1 dimensions is developed. The approach is
based on the similarity between growing braid and  "heap of colored pieces". As
an application, the problem of a vortex glass transition in high--$T_{\rm c}$
superconductors is re-examined on microscopic level.
\end{abstract}
\medskip

Statistics of ensembles of uncrossible linear objects with topological
constraints has very broad application area ranging from problems of
self--diffusion of directed polymer chains in flows and nematic--like
textures to dynamical and topological aspects of vortex glasses in
high temperature superconductors \cite{nel}. In this letter we
propose a microscopic approach to a diffusive dynamics of entangled
uncrossible lines of arbitrary physical nature.

It is well known that the main difficulties in the statistical topology of
linear uncrossible objects are due to two facts: (a) the topological
constraints are non--local and (b) different entanglements do not commute.
The point (a) can be overcomed by introducing the corresponding (abelian)
Gauss--like topological invariant which properly counts windings of one
chain around the other, while the circumstance (b) since now creates the
major problem in constructive approach to topological theories beyond
the abelian approximation. In order to have representative and physically
clear image for the system of fluctuating lines with nonabelian topology
we formulate the model in terms of entangled Brownian trajectories: such
representation serves also as a geometrically clear image of Wilson loops
in (2+1)D nonabelian field--theoretic path integral formalism.

We are aimed to develop a symbolic language which would permit us to construct
the objects with a braid--like topology in 2+1 dimension and to solve the
simplest statistical problems where the noncommutative character of topological
constraints is properly taken into account. The results are applied to
re-examination of the problem of a vortex glass transition in high--$T_{\rm c}$
superconductors \cite{ob_rub}. Let us remind briefly that in ${\rm
Cu0_2}$--based high--$T_{\rm c}$ superconductors in fields less than $H_{\rm
c2}$ there exists a region where the Abrikosov flux lattice is molten, but the
sample of the supereconductor demonstrates the absence of the conductivity.
This effect is explained by highly entangled state of flux lines due to their
topological constraints \cite{ob_rub}.

The most attention in our investigation is paid to a quantitative
estimation of a characteristic time of self--disentanglement of a
particular "test" chain in a bunch of braided directed chains. We
distinguish between two situations: (i) configurations of all lines in a
bunch are quenched and form a lattice except one test line randomly
entangled with the others, and (ii) no chain in a bunch of braided
lines is fixed and any chain winds randomly like a test one. We compute
the characteristic times of topological relaxations $ \tau_{\rm si}$ and
$\tau_{\rm co}$ in cases (i) and (ii) and demonstrate the {\it absence of
qualitative difference} between $\tau_{\rm si}$ and $\tau_{\rm co}$,
however the quantitative distinction between $\tau_{\rm si}$ and $\tau_{\rm
co}$ is shown to be sufficiently  strong. Our result contradicts in details
with the statement of \cite{ob_rub} on {\it qualitative difference} between
$\tau_{\rm si}$ and $\tau_{\rm co}$ obtained in the frameworks of a scaling
analysis. Nevertheless our result does not destroy the physical conclusions
of the paper \cite{ob_rub} about the possibility of topological glass
transition in entangled flux state in high--$T_{\rm c}$ superconductors.
According to the above mentioned cases (i) and (ii) we define respectively
two discrete models I and II.

\begin{figure}
\begin{center}
\epsfig{file=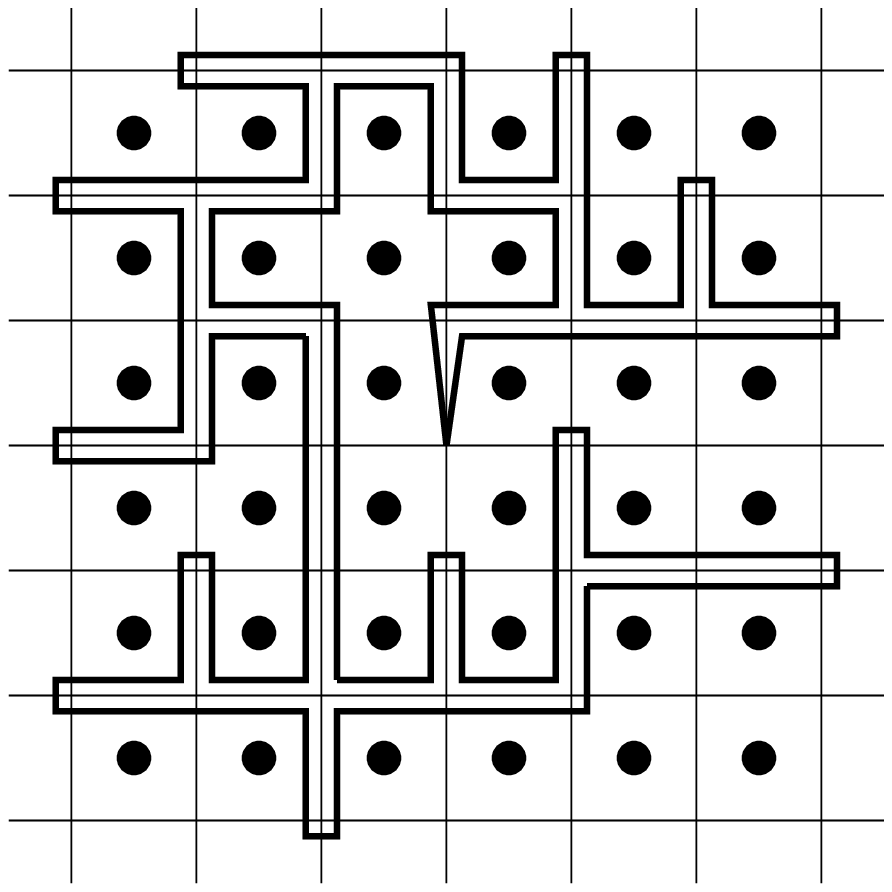,width=3.8cm}
\end{center}
\caption{The $(xy)$--projection of a path unentangled with the square
lattice of topological obstacles.}
\label{fig:1}
\end{figure}

The {\it model I} is as follows. Take a square lattice in $(xy)$--plane
with a spacing $c$ and put in all vertices of this lattice the uncrossible
obstacles. Consider a symmetric random walk with the step length $c$ on a
dual lattice shifted by $\frac{c}{2}$ in both  $x$-- and $y$--directions.
We are interested in computing the probability $P_{\rm si}(N)$ of the fact
that after $N$ steps on the dual lattice the random path will be closed
and unentangled with respect to the obstacles. It is clear that in the
(2+1)--dimensional "space--time" ${\Z}^2(x,y) \times {\Z}^{+}(t)$ this
model describes statistics of a "world  lines" (time--ordered paths) of a
single particle jumping on a square lattice  in $(xy)$--projection, making
each time a step toward $t$--axis and topologically interacting with the
lattice of infinitely long straight lines. The $(xy)$--section of this
model is shown in fig.\ref{fig:1}.

It is well known that there exists a bijection between a path in the lattice of
obstacles with coordinational number $z$ and a path on a Cayley tree with $z$
branches \cite{kh_ne}. Different topological states of a given path coincide
with the elements of the homotopy group of the multi--punctured plane
$\Gamma_{\infty}$, generated by a countable set of elements. The translational
invariance allows to consider in a local basis the group
$\Gamma_{\infty}/{\Z}^2= \Gamma_{z/2}$, where $\Gamma_{z/2}$ is a free group
with $z/2$ generators i.e. a $z$--branching tree. Any topological state of a
path in the lattice of obstacles is encoded by an element of the group
$\Gamma_{z/2}$, and uniquely corresponds to some irreducible word written in
terms of generators of the group $\Gamma_{z/2}$. The length of this irreducible
word coincides with the geodesic distance along the Cayley graph of the group 
$\Gamma_{z/2}$. The average "degree of entanglement" of an $N$--step path in
the lattice of obstacles with coordinational number $z$ for $N\gg 1$ is
characterized by the averaged geodesic distance \cite{kh_ne} $\left<L_{\rm
si}(N)\right>=\frac{z-2}{z}N$ on the Cayley tree with $z$ branches.  Hence, the
normalized "complexity" $\left<l_{\rm si}\right>$ of a  typical topological
state of a path can be defined as:
\be \label{eq:0}
\left<l_{\rm si}\right>\equiv
\lim_{N\to\infty}\frac{\left<L_{\rm si}(N)\right>}{N}=\frac{z-2}{z}
\ee

For diffusion on the Cayley tree, $P_{\rm si}(k,N)$ deftermines the probability
for the $N$--step symmetric random walk to have a distance between ends in $k$
steps along the tree. This probability has been computed many times---see, for
example, \cite{kes}. Thus we reproduce the final result for the return
probability $P_{\rm si}(k=0,N)$ as $N\to\infty$ on the Cayley tree:
\be \label{eq:1}
P_{\rm si}(N)\equiv P_{\rm si}(k=0,N)=\frac{2\sqrt{2}p}{\sqrt{\pi}(1-4pq)}
\frac{\alpha^N}{N^{3/2}}
\ee
where $q=\frac{1}{z}$ and $p\equiv 1-q=\frac{z-1}{z}$ are the probabilities  of
steps "towards" and "backwards" the origin of the Cayley tree;
$\alpha=2\sqrt{pq}$. Define $\Lambda$, a span of the $N$--step random walk in
$(xy)$--projection. In physical terms of the original "vortex problem",
$\sqrt{\Lambda^2}$ is the averaged size of thermal fluctuations of the vortex
line. Hence, we can set $N=\frac{\Lambda^2}{c^2}$ and estimate the  time of
topological relaxation $\tau_{\rm si}=\frac{1}{P_{\rm si}(N)}$ of a single
vortex line in an ensemble of immobile uncrossible lines for $z=4$ as follows
\be \label{eq:2}
\tau_{\rm si}\sim\left(\frac{\Lambda^2}{c^2}\right)^{3/2}
\left(\frac{2}{\sqrt 3}\right)^{\frac{\Lambda^2}{c^2}}
\ee

\begin{figure}
\begin{center}
\epsfig{file=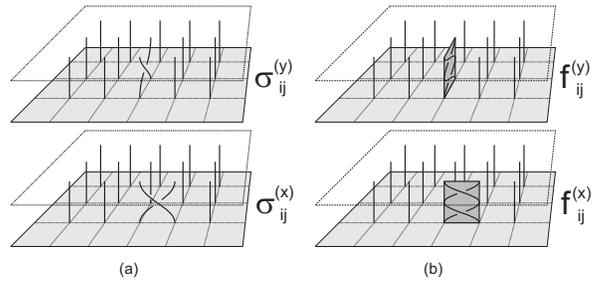,width=2.4cm}
\end{center}
\medskip

\caption{The generators of the surface groups $B_n^{2D}$ (a) and ${\cal
LF}_{n+1}^{2D}$ (b).}
\label{fig:2}
\end{figure}

\begin{figure}
\begin{center}
\epsfig{file=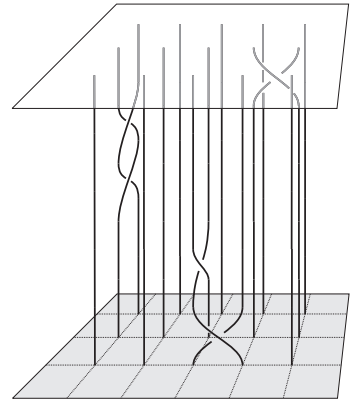,width=3.6cm}
\end{center}
\caption{The (2+1)--dimensional braid.}
\label{fig:3}
\end{figure}

In contrast to the model I, the {\it model II} describes collective
dynamics of the world lines and ultimately leads to the consideration of
the (2+1)--dimensional ("surface") {\it braid group} $B_{n+1}^{2D}$.
Consider the two--dimensional lattice ${\Z}^2(x,y)$ and take distinct
points $P_1, P_2, ...P_{(n+1)^2}, \in {\Z}^2$.
A (2+1)--braid of $(n+1)^2$ strings on ${\Z}^2$ based at $\{P_1, ...,
P_{(n+1)^2}\}$ is an $(n+1)^2$--tuple $b=(b_1,\ldots, b_{(n+1)^2})$ of paths,
such that: (1) $b_i(1)=P_i$ and $b_1(1)\in \{P_1, ..., P_{n^2}\}$ $\forall i\in
\{1, ..., (n+1)^2\}$; (2) $b_i(t)\neq b_j(t)$ $\forall \{i,j\}\in \{1, ...,
(n+1)^2\}$, $(i\neq j,\; t\in [1,N])$. The braid group $B^{2D}_{n+1}$ on
${\Z}^2$ based at $\{P_1, ..., P_{(n+1)^2}\}$ is the group of homotopy classes
of braids based at $\{P_1, ..., P_{(n+1)^2}\}$. The group $B^{2D}_{n+1}$ has
$2n^2+2n$ generators $\sigma_{ij}^{(x)}$, $i\in \{1,\ldots,n+1\}, j\in
\{1,\ldots,n\}$ ; $\sigma_{ij}^{(y)}$, $i\in \{1,\ldots,n\}, j\in
\{1,\ldots,n+1\}$ and their inverses with the standard relations \cite{bir}.
The geometric representation of generatoes of $B^{2D}_{n+1}$ is shown in
fig.\ref{fig:2}a. An element of the braid group $B^{2D}_{n+1}$ is set  by a
word in the alphabet $\{\sigma_{11}^{(x)}, \sigma_{11}^{(y)}, \ldots\}$. By the
{\it length} $N$ of a braid we call a length of a word in a given record of the
braid, and by the irreducible length (or {\it primitive length})---the minimal
length of a word, in which the braid can be written. The irreducible
length can be also viewed as a distance from the unity on the graph of the
group. Graphically the braid is represented by a set of strings, going upwards
in accordance with a growth of a braid length---see fig.\ref{fig:3}.

Define now a symmetric random walk on a set of generators ("letters")
$\{\sigma_{11}^{(x)}, \sigma_{11}^{(y)}, \ldots\}$ with the transitional
probability $\frac{1}{2n^2+2n}$. Namely, we raise recursively an $N$--letter
random word $W$ adding step-by-step the letters (say, from the right--hand
side) to a growing word. The probability that the $N$--letter word is
completely contractible (i.e. has a zero's  primitive length) defines the
probability to have topologically trivial braid of the record length $N$.

\begin{figure}
\begin{center}
\epsfig{file=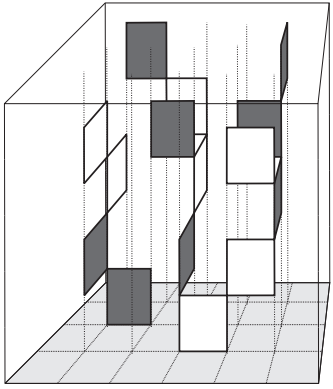,width=3.6cm}
\end{center}
\caption{The (2+1)--dimensional colored heap.}
\label{fig:4}
\end{figure}

Our main tool in the investigation of the braid group is the so-called {\it
locally free group} \cite{vnb}. The (2+1)D ("surface") locally free group
${\cal LF}_{n}^{2D}$ is obtained from the braid group $B_{n+1}^{2D}$ by
omitting the braiding relations. The group ${\cal LF}^{2D}_n$ has
$2n^2+2n$ generators $f_{ij}^{(x)}$, $i\in \{1,\ldots,n+1\}, j\in
\{1,\ldots,n\}$ ; $f_{ij}^{(y)}$, $i\in \{1,\ldots,n\}, j\in
\{1,\ldots,n+1\}$   and their inverses with the
relations (see also fig.\ref{fig:2}b):
$$
\left\{\begin{array}{l}
f_{i_1,j_1}^{(x)}f_{i_2,j_2}^{(x)}=
f_{i_2,j_2}^{(x)}f_{i_1,j_1}^{(x)}\;
(|j_1-j_2|>0\; {\rm or}\; |i_1-i_2|>1) \\
f_{i_1,j_1}^{(x)}f_{i_2,j_2}^{(y)}=
f_{i_2,j_2}^{(y)}f_{i_1,j_1}^{(x)}\;
(i_2-i_1 \; {\rm or}\; j_1-j_2) \ne \{0,1\}) \\
f_{i,j}^{(x)}\left(f_{i,j}^{(x)})\right)^{-1}=
f_{i,j}^{(y)}\left(f_{i,j}^{(y)})\right)^{-1}=e
\end{array} \right.
$$
There is a bijection between words in locally free group and {\it
colored heaps}, whose elements are either "white" $f_{ij}^{(x,y)}$ or
"black" $\left(f_{ij}^{(x,y)}\right)^{-1}$. That is, any word written in
terms of letters--generators of the group  ${\cal LF}^{2D}_n$ represents
a configuration of a colored heap (see fig.\ref{fig:4}) in a box of the
base of $n\times n$ cells and any such heap uniquely defines some word in
the group ${\cal LF}^{2D}_n$. The configuration of the heap with a
"black" block following immediately after a "white" one in the same column
is forbidden.

For the uniform Markov dynamics on the set of generators we can compute the
average primitive length $\left<L_{\rm co}(N)\right>$ of the $N$--letter word
which characterizes the degree of entanglement of threads in a braid (compare
to the definition of $\left<L_{\rm si}(N)\right>$). Let us take into account
that ${\cal LF}^{2D}_n$ is a subgroup of $B^{2D}_{n+1}$, while $B^{2D}_{n+1}$
is a factor--group of ${\cal LF}^{2D}_n$. This is manifested in the two
following facts. (1) By definition the locally free group ${\cal LF}^{2D}_n$
has less relations than the braid group $B^{2D}_{n+1}$. Hence, the number of
distinct words of primitive length $L_{\rm co}$ in braid group {\it is bounded
from above} by the number of distinct words of the same primitive length
$L_{\rm co}$ in the locally free group. (2) By construction (compare
figs.\ref{fig:2}a and \ref{fig:2}b) $f^{(x,y)}_{i,j}=\left(\sigma^{(x,y)}_{i,j}
\right)^2  \quad (\{i,j\}\in[1,n])$. Thus, the number of distinct words of
length $2L_{\rm co}$ in the braid  group {\it is bounded from below} by the
number of distinct words of the length  $L_{\rm co}$ in the locally free group
\cite{vnb}. The facts (1)--(2) allow us to get the bilateral estimation for the
average primitive length $\left<L_{\rm co}(N|B^{2D}_{n+1})\right>$ of the
$N$--step random walk on the surface braid group $B^{2D}_{n+1}$:
\be \label{eq:3}
\frac{1}{2}\left<L_{\rm co}(N|{\cal LF}^{2D}_n)\right>\le
\left<L_{\rm co}(N|B^{2D}_{n+1})\right>\le
\left<L_{\rm co}(N|{\cal LF}^{2D}_n)\right>
\ee

\begin{figure}
\begin{center}
\epsfig{file=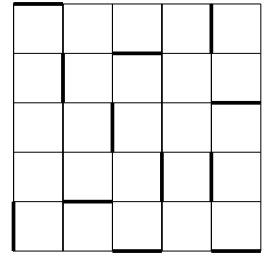,width=3.4cm}
\end{center}
\caption{A particular configuration of a roof of (2+1)D heap is shown by
bold segments.}
\label{fig:5}
\end{figure}

The computation of $\left<L_{\rm co}(N|{\cal LF}^{2D}_n)\right>$ involves
the concept of the {\it roof of the heap}. In physical terms a roof of a
heap consists of a set of "most top" blocks which can be removed from the
heap without changing the rest of it. The projection of a roof to the
$(xy)$--plane for some particular configuration of the most top blocks is
shown in fig.\ref{fig:5}. Let us stress that local hights (measured from the
bottom of the box) of different roof's blocks might be different.

The process of growth of a heap (i.e. the random walk on the surface locally
free group ${\cal LF}^{2D}_n$) consists in adding step-by-step new "black" or
"white" blocks to the roof. Hence the dynamics of a heap is controlled by the
dynamics of a roof. For a particular configuration of a roof we define the
"size" of a roof $\#T$ (i.e. the number of bold segments in fig.\ref{fig:5})
and the number of empty segments $n_i$ having $i$ bold neighbors (apparently,
$n_i=0\, \forall\, i\ge 3$). In fig.\ref{fig:5} one has $n=5,\, n_0=6,\,
n_1=25,\, n_2=17;\; \#T=12$. For the values $\#T,\,n_i$ the following
conditions hold:
\be \label{eq:4}
\left\{\begin{array}{l}
n_0+n_1+n_2+\#T=2n^2+2n \\
6\#T-8(n+1)\le n_1+2n_2\le 6\#T
\end{array}\right.
\ee
For a given configuration the local dynamics of a size of a roof reads
$$
\left\{\begin{array}{ll}
\Delta \#T=1 & \mbox {with probability $\frac{n_0}{2n^2+2n}$} \\
\Delta \#T=0 & \mbox {with probability $\frac{n_1+\#T}{2n^2+2n}$} \\
\Delta \#T=-1 & \mbox {with probability $\frac{n_2}{2n^2+2n}$}
\end{array}\right.
$$
which together with (\ref{eq:4}) allows to estimate $\#T$ as follows
$$
1-\left<\Delta \#T\right>\le\frac{7\#T}{2n^2+2n}\le 1+\frac{8(n+1)}{2n^2+2n}-
\left<\Delta \#T\right>
$$
In a stationary case we have $\left<\Delta \#T\right>=0$, what permits to
get the asymptotic value $\left<\#T\right>$  of the average number of
block in the roof for $n\gg 1$:
\be \label{eq:7}
\left<\#T\right>=\frac{2n^2}{7}
\ee

Return to the random walk on ${\cal LF}^{2D}_n$. The conditional
change of the primitive length $L_{\rm co}(N|{\cal LF}^{2D}_n)$ for one step
of the random walk is
$$
\left\{\begin{array}{ll}
\Delta L_{\rm co}(N|{\cal LF}^{2D}_n)=1 & \mbox {with probability
$1-\frac{\#T}{2(2n^2+2n)}$} \\
\Delta L_{\rm co}(N|{\cal LF}^{2D}_n)=-1 & \mbox {with probability
$\frac{\#T}{2(2n^2+2n)}$}
\end{array}\right.
$$
Taking into account (\ref{eq:7}) we get in a stationary state (when $N\gg
1$ and $n\gg 1$) the following asymptotic value for the average primitive
length $\left<L_{\rm co}(N|{\cal LF}^{2D}_n)\right>$:
$$
\left<L_{\rm co}(N|{\cal LF}^{2D}_n)\right>=\frac{6}{7}N
$$
Thus, according to (\ref{eq:3}) we arrive at the bilateral
estimation for the average length of the primitive word for the $N$--step
random walk on the surface braid group
\be \label{eq:9}
\frac{3}{7}\le \lim_{N\to\infty \atop n\to\infty}
\frac{\left<L_{\rm co}(N|B^{2D}_n)\right>}{N}
\le\frac{6}{7}
\ee

The quantity $\left<l_{\rm co}\right>=\lim\limits_{N\to\infty \atop n\to\infty}
\frac{\left<L_{\rm co}(N|B^{2D}_n)\right>}{N}$ characterizes the "complexity"
of entangled state. Comparing (\ref{eq:9}) to (\ref{eq:0}) we can conclude that
the averaged topological state of a braid $\left<L_{\rm co}(N|B^{2D}_n)\right>$
obtained in course of a collective motion of lines has the same asymptotics in
$N$ as the one of a single line motion and interpolates between entangled
states in effective lattices of obstacles with coordinational numbers $z_{\rm
eff}$ ranging in the interval $\left[\frac{7}{2},14\right]$.

Let us utilize now the "heap concept" and eq.(\ref{eq:7}) for estimating
{\it from above} the time of topological relaxation in a bunch of vortex
lines, considered as a braid of directed random walks. The random growth
of a braid of $2n^2+2n$ lines of length $N$ each, can be interpreted as an
$N$--step random walk on a surface braid group $B^{2D}_{n+1}$. The
topological state of a braid is uniquely characterized by a primitive word
$W_N$ in terms of generators of $B^{2D}_{n+1}$. The disentangled state of
two neighboring trajectories means (in terms of the group $B^{2D}_{n+1}$)
that the primitive word $W_N$ does not contain the generators
$\sigma^{(x)}_{i,j}, \sigma^{(y)}_{i,j},
\left(\sigma^{(x)}_{i,j}\right)^{-1},
\left(\sigma^{(x)}_{i,j}\right)^{-1}$ for some $1\le \{i,j\}\le n$.

Let $L_{\sigma}(W_N)$ be the number of generators $\sigma=
\left\{\sigma^{(x)}_{i,j},\left(\sigma^{(x)}_{i,j}\right)^{-1}\right\}$ in
the primitive word $W_N$. Then the function $P_{\rm co}\{x,N\}\equiv
P\{L_{\sigma}(W_N)=x\}$ defines the characteristic time
$\tau_{\rm co}=\frac{1}{P_{\rm co}(0,N)}$ of disentanglement of a
particular line in course of collective Brownian motion of all $2n^2+2n$
lines in a braid.  Suppose that after $N$ random steps we arrive at the word
$W_N$. Then at the step $N+1$ we have $W_{N+1}=W_N\, W_{NN+1}$. The
probability that the word $W_{NN+1}$ contains a generator from the set
$\sigma$ (with a prescribed sign) is $P\{\sigma\in
W_{NN+1}\}=\frac{1}{4}\times\frac{1}{8}=\frac{1}{32}$. Consider the
probability $q^b_N=P\{L_{\sigma}(W_{N+1})= L_{\sigma}(W_{N+1})-1\}$
of reducing the element $\sigma$ by $1$ at the step $N+1$. The probability
$q^b_N$ we can estimate from below, replacing the word $W_{N+1}$ by the word
$\tilde{W}_{N+1}$ which itself is obtained by replacing the generators of the
group $B^{2D}_{n+1}$ by the ones of the group ${\cal LF}^{2D}_n$. The word
$\tilde{W}_{N+1}$ is also primitive by construction and
$q^l_N=P\{L_f(\tilde{W}_{N+1})= L_f(\tilde{W}_{N+1}-1)\}
\le q^b_N$, where $f\in \left\{f^{(x)}_{i,j},
\left(f^{(x)}_{i,j}\right)^{-1}\right\}$. Denote by $Q^l_N$ the
probability that after $N$ steps the roof contains the generators
$f^{(x)}_{ij}$ or $f^{(y)}_{ij}$. In the case $n>>1$ we can neglect the
'boundary effects' caused by marginal generators  $f_{n+1,j}^{(x)}$ , $f_{1,j}^{(x)}$
, $j\in\{1,\ldots,n\}$ and $f_{i,n+1}^{(y)}$ , $f_{i,1}^{(y)}$
, $i\in\{1,\ldots,n\}$ so that the probability $Q^l_N$ is the same for all
$f\in \left\{f^{(x)}_{i,j},
\left(f^{(x)}_{i,j}\right)^{-1}\right\}$. Thus we have the obvious
expression for the average number of blocks in the roof in the stationary
case $N\gg1$ :  $\left<\#T\right>=2n^2Q^l$; which gives $Q^l=\frac17$.
 Then the probability of cancellation of
the generator $f$ at the step $N+1$ is $q^l_N=\frac{1}{32}Q^l_N= \frac{1}{224}\le q^b$
where $q^b$ is the stationary probability of cancellation of generator $\sigma$
at $N\to\infty$. Substituting $q^b$ for $q$ in  (\ref{eq:1}) we get the
estimate from above for the time $\tau^{\rm up}_{\rm co}$:
\be \label{eq:10}
\tau^{\rm up}_{\rm co}\sim\left(\frac{\Lambda^2}{c^2}\right)^{3/2}
\left(\frac{112}{\sqrt{223}}\right)^{\frac{\Lambda^2}{c^2}}
\ee
The numerical analysis of statistics of braid and locally free groups
\cite{dub_ne} enables us to conjecture that the estimate $\tau^{\rm
up}_{\rm co}$ for the group ${\cal LF}^{2D}_n$ is close to the
value $\tau_{\rm co}$ for the group $B^{2D}_n$.

Comparing (\ref{eq:10}) to (\ref{eq:2}) we conclude that the scaling
dependences of the characteristic times $\tau_{\rm si}$ and $\tau_{\rm
co}$ on $\frac{\Lambda}{c}$ are the same, however numerically $\tau_{\rm
co}$ is much larger than $\tau_{\rm si}$, namely
$\lim\limits_{\frac{\Lambda}{c}\to\infty}\frac{c^2}{\Lambda^2} \ln
\frac{\tau_{\rm co}}{\tau_{\rm si}}\approx 6.5$.

Summarizing the said above, let us emphasize that the random braiding  can be
analyzed within the frameworks of a symbolic dynamics on the locally  free
group describing the growth of a heap of colored pieces. The probability to
have disentangled state of two vortex lines can be estimated from above by  the
probability to have no pieces (blocks) in a given column of a heap. This model
seems to be a natural discretization of a standard ballistic growth process of
Kardar--Parisi--Zhang type \cite{kpz}.

The authors are grateful to A.Vershik and J.Desbois for useful discussions.

\end{document}